# Anti-aliasing Algorithm Based on Three-dimensional Display Image

Ziyang Liu 12211411 Xingchen Xiao 12212904 Yueyang Xu 12111106

***Abstract*—3D-display technology has been a promising emerging area with potential to be the core of next-generation display technology. When directly observing unprocessed images and text through a naked-eye 3D display device, severe distortion and jaggedness will be displayed, which will make the display effect much worse. In this work, we try to settle down such degradation with spatial and frequency processing, furthermore, we make efforts to extract degenerate function of columnar lens array thus fundamentally eliminating degradation.**

## I. INTRODUCTION

Cylindrical lenses, widely used in three-dimensional (3D) display technology, possess unique optical properties that can lead to significant image quality issues. Among these issues, aliasing effects are particularly prominent, caused by the loss of detail in low-resolution or high-contrast regions. These effects manifest as jagged and unsmooth edges, especially at specific angles, which reduce the overall visual quality of the display. Such distortions are exacerbated when observing unprocessed images and text through naked-eye 3D display devices, resulting in severe visual artifacts and compromised user experience. [1-2]

The tilt angle of the cylindrical lens array is a critical factor in this distortion. It has been observed that the most significant aliasing occurs at an angle of approximately 80 degrees, corresponding to the tilt of the lens. The interference caused by the cylindrical lens introduces additional nodes to otherwise smooth straight lines, further complicating image clarity. This phenomenon highlights the necessity for an effective anti-aliasing algorithm to mitigate these distortions.

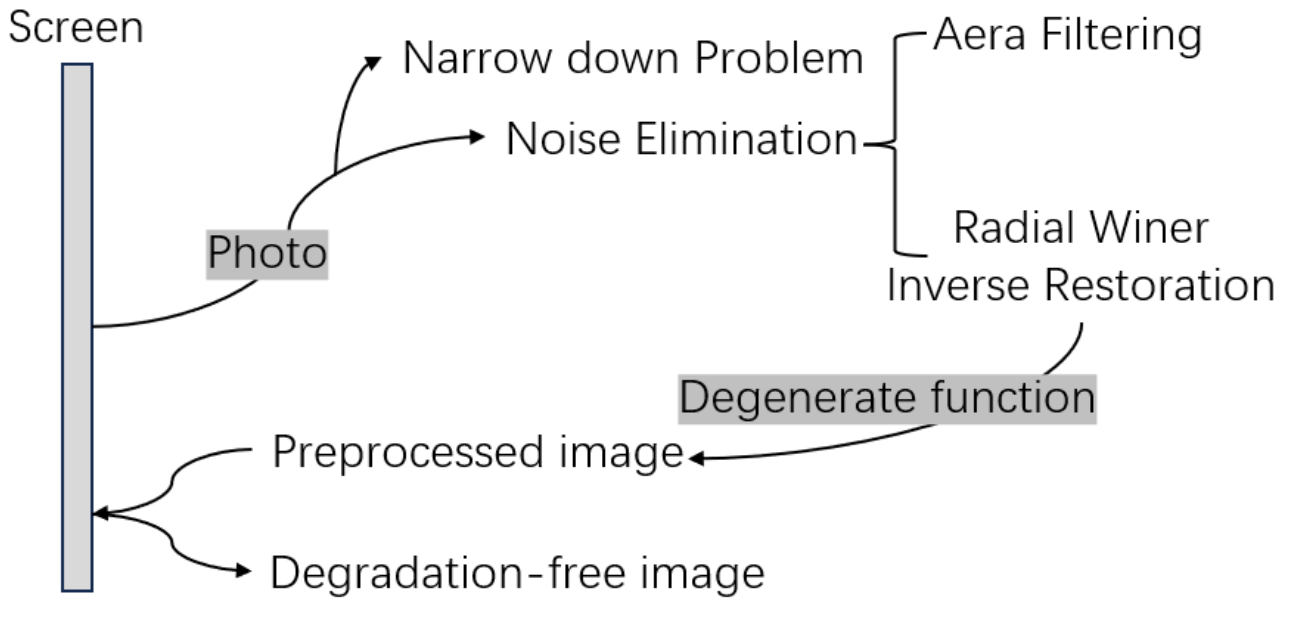

Figure 1 Diagram of process flow

Thus, it is crucial to consider the degradation phenomenon, for it might obstruct higher-resolution displays and their commercialization. In this work, we narrow down the degradation to the tilt angle of the lens array and propose a fully innovative process flow to eliminate it. The further analysis of the degenerate function might contribute to the understanding of the optical mechanism behind the degradation and help to solve it at the hardware level.

Because the degradation is due to the lens array. We can only take photos to obtain the degenerated diagram. We firstly use a series of pictures to narrow down the tilt angle and pattern that have prominent distortion. After obtaining the worst case, we apply the global threshold and multiple median filters to get binary diagram. Then we have tried two branches of solutions to the degenerated diagram: direct frequency-domain filtering and inverse restoration.

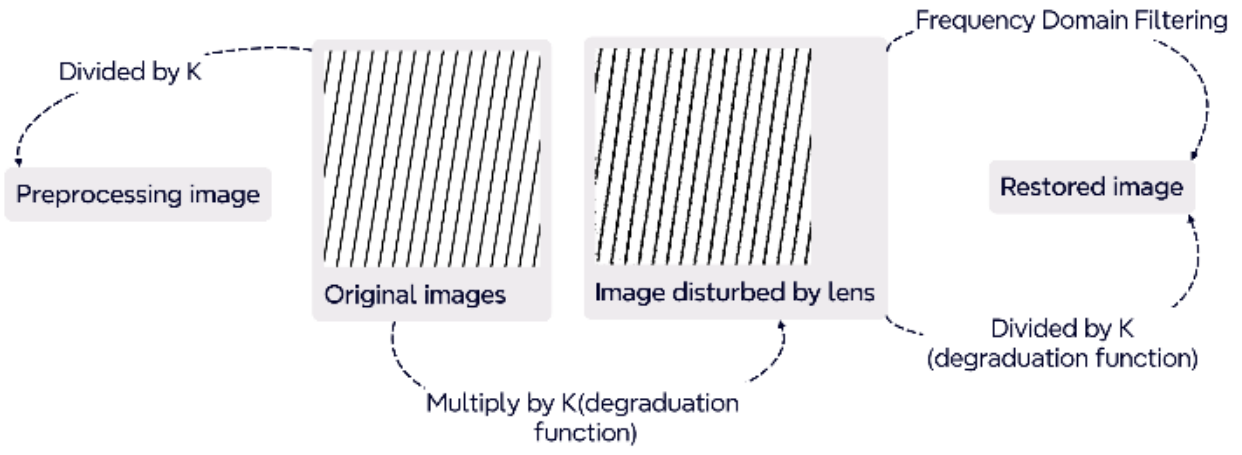

Figure 1 Process flow of inverse restoration

Given that the distortion is spatially periodic (if you only consider "nodes" appearing in equidistance lines), we can infer from the space-frequency domain symmetrical property that the degradation function H is not a periodic or symmetrical function in the frequency domain. By comparing the frequency domain of degraded and nondegraded diagrams, we have proposed several frequency-domain filtering methods.

## II. RESTATEMENT OF PROBLEM

To show the distortion and aliasing of a line along different directions, circular pattern with different tilt line had been drawn.

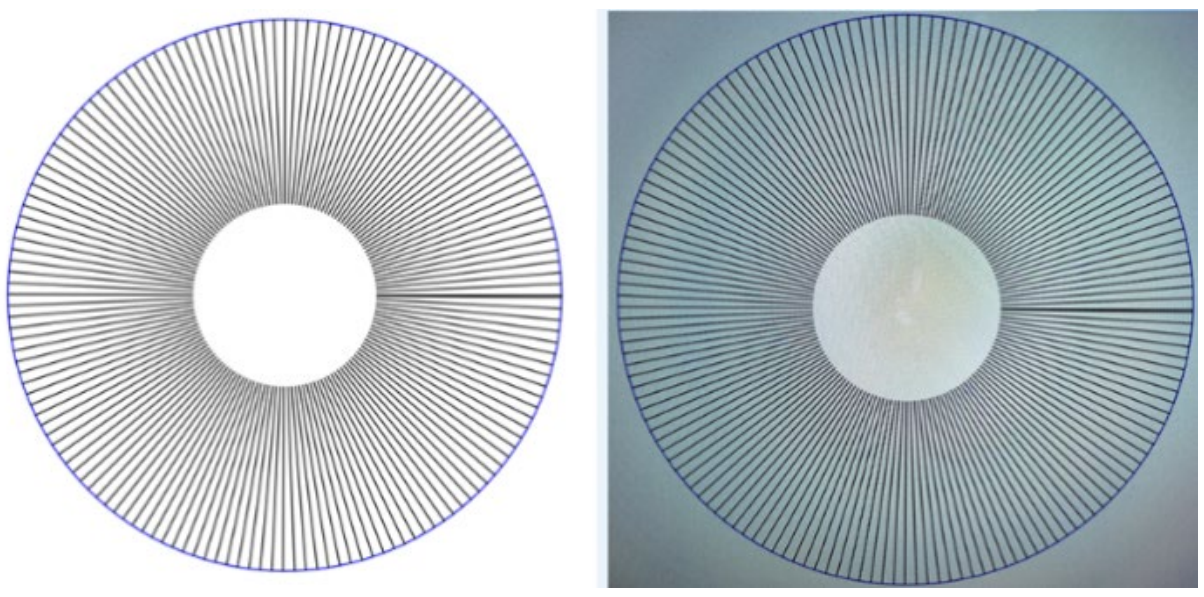

Figure 2 Ideal (a) and photo (b) of circular pattern with varying tilt

As can be seen from this figure, the angle with the most severe distortion is about 80 degrees.

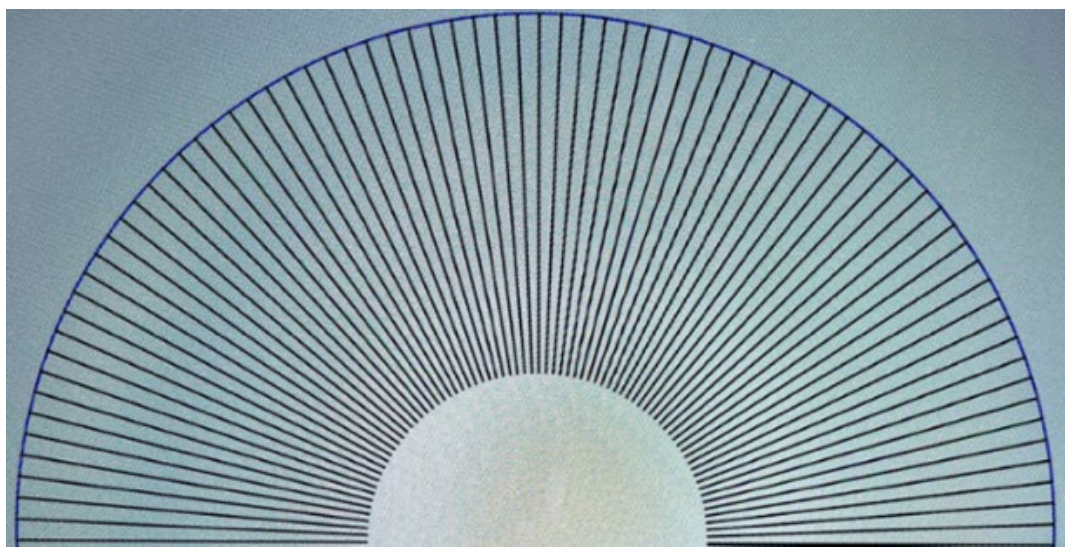

Figure 3 Enlarged circular pattern with varying tilt

The tilt angle of the cylindrical lens is also about 80 degrees, so we conclude that the most serious distortion is along the tilt direction of the cylindrical lens.

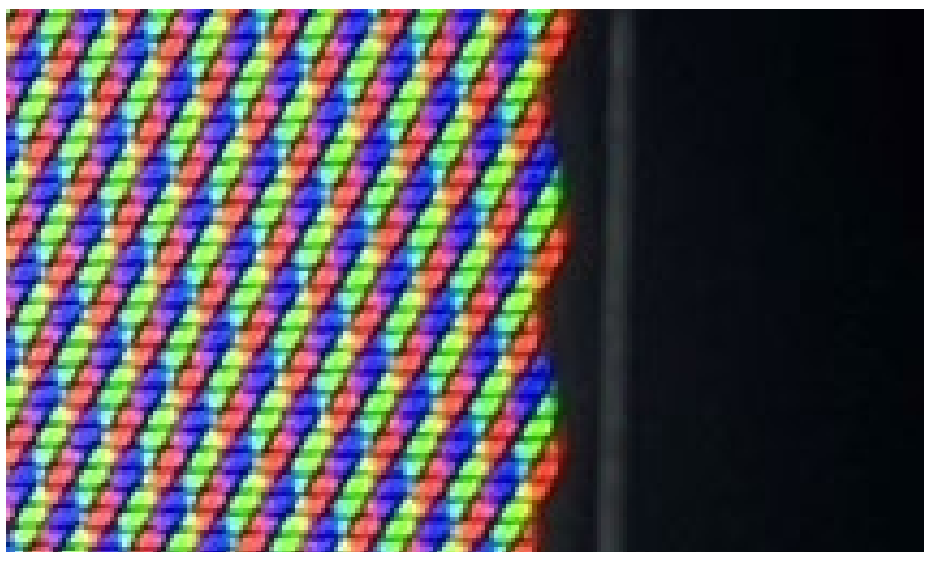

Figure 4 Photos of pixel observed through lens array

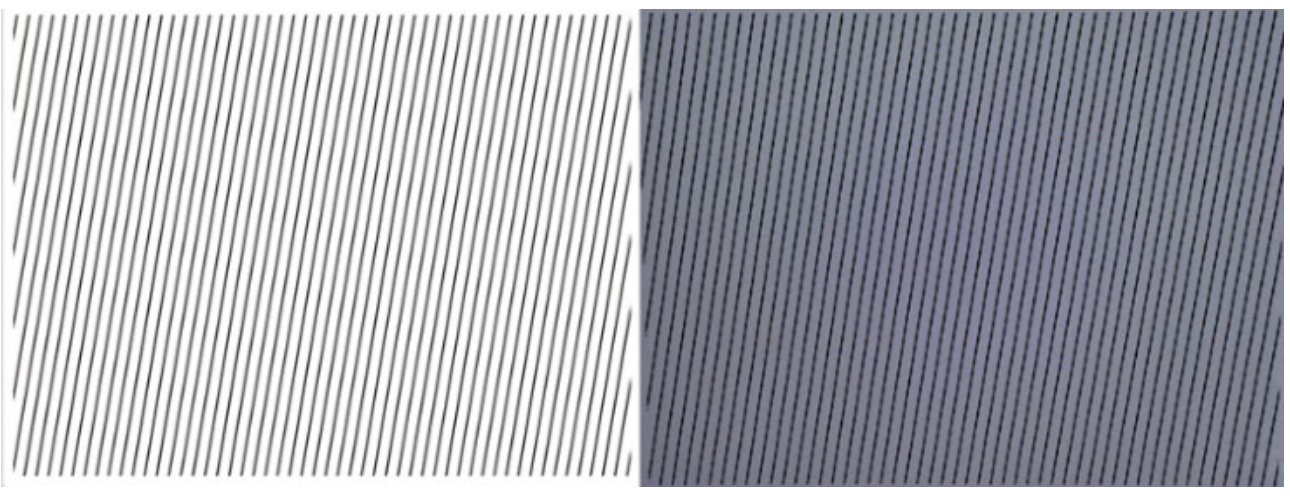

Figure 5 Ideal (a) and actual photo (b) of Equidistance lines with 80-degree tilt angle

In order to better observe the aliasing and distortion along the 80-degree line, straight lines with an angle of about 80 degrees are drawn.

## III. Image preprocessing

The histogram in Fig.7.(b) presents two peaks in the grey value, which correspond to the useful patterns. The values between 20 to 230 correspond to the noise introduced by pixels and photo taking.

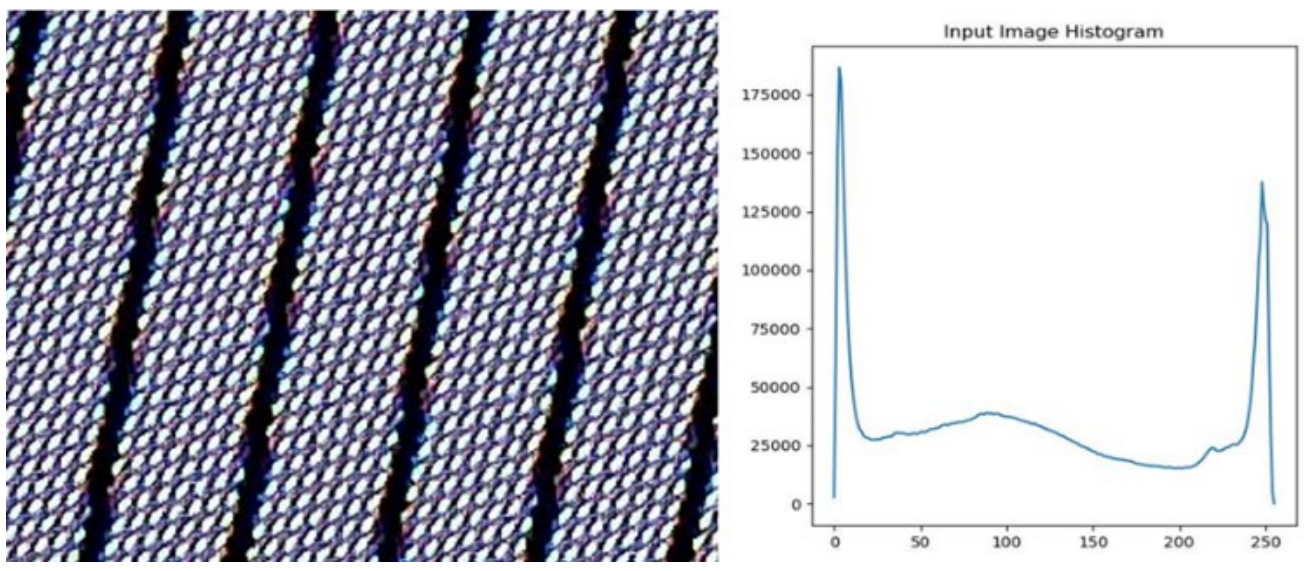

Figure 6 (a) Amplified screen-shot and (b) corresponding histogram

Given that the high-frequency noise introduced by the fringe of pixels will affect our restoration. We need to apply a global threshold filtering. After the pre-test we found that due to the limitation in photographing, pixel values around the "nodes" get close to the grey values of noise, making it impossible for the direct application of global thresholding filtering.

To filter out noise while maintaining the node pattern, we use a median filter next to a global thresholding filtering to acquire a high-quality diagram with clear nodes and eliminate noise caused by photo taking. After several attempts, we obtain the best filtering outcomes (in Fig.8) by using $T = 60$ and multiple median filters with window size = $7 * 7$ .

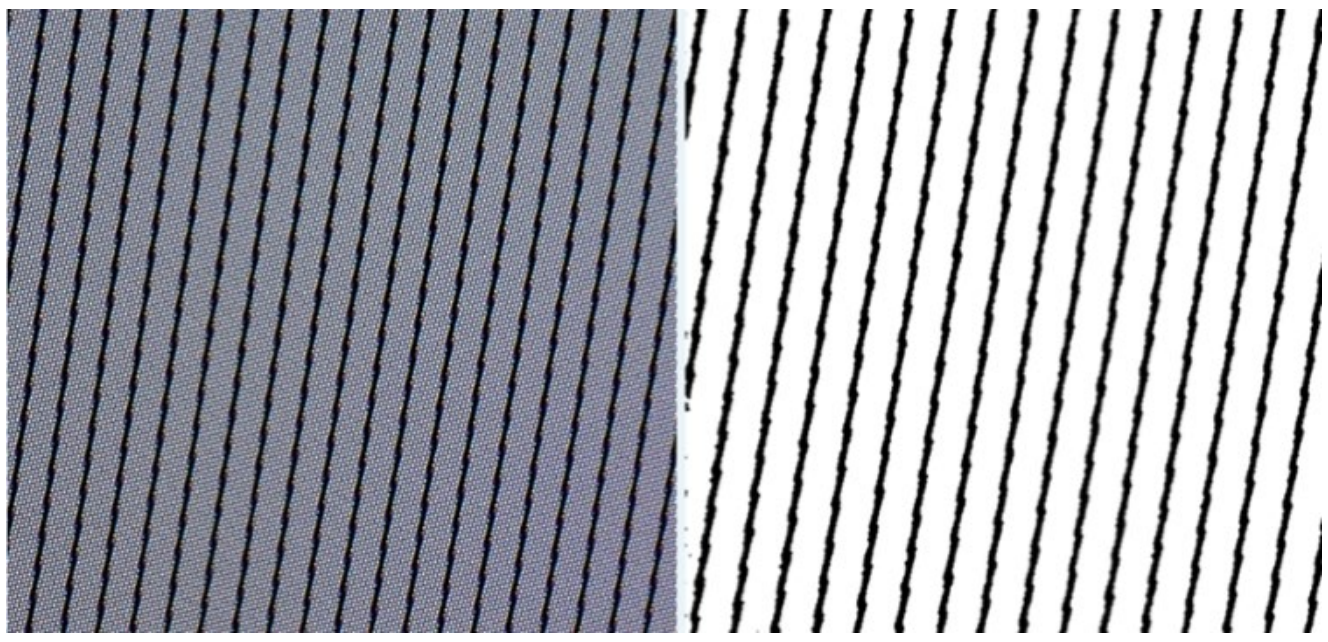

Figure 8 Diagram before (a) and after (b) binary-median processing

## III. Frequency-domain filtering

The frequency-domain filtering method leverages frequency-domain information to identify and remove aliasing artifacts caused by cylindrical lenses in three-dimensional display systems. By applying the Fourier Transform, the image is converted from the spatial domain to the frequency domain, where aliasing features appear as distinct lines in the magnitude spectrum, making them easier to locate and address.

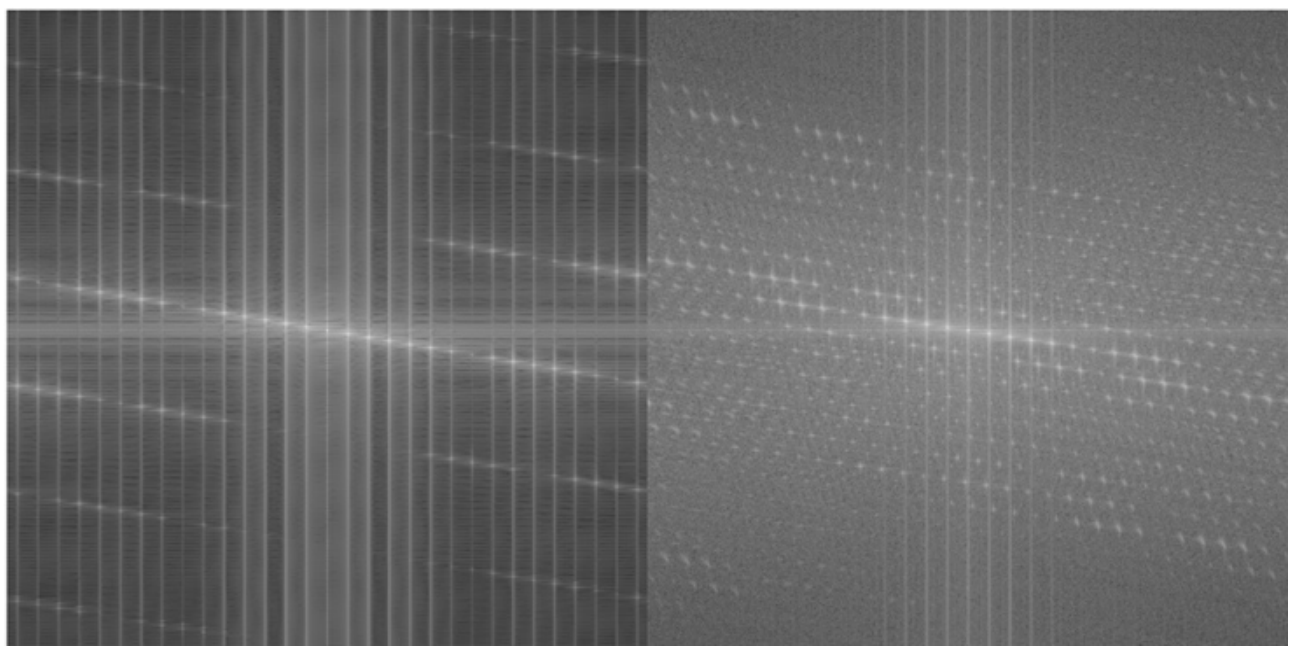

Figure 7 Frequency domain of original (a) and degraded (b) diagram

In identifying the aliasing direction, distortions that occur at approximately 80 degrees in the spatial domain are represented as lines at around 170 degrees in the frequency domain. The reason for this phenomenon is that the line exhibits no change in the direction of 80 degrees, yet undergoes an abrupt change at 170 degrees. Consequently, the line's direction in the frequency domain is perpendicular to its direction in the spatial domain.

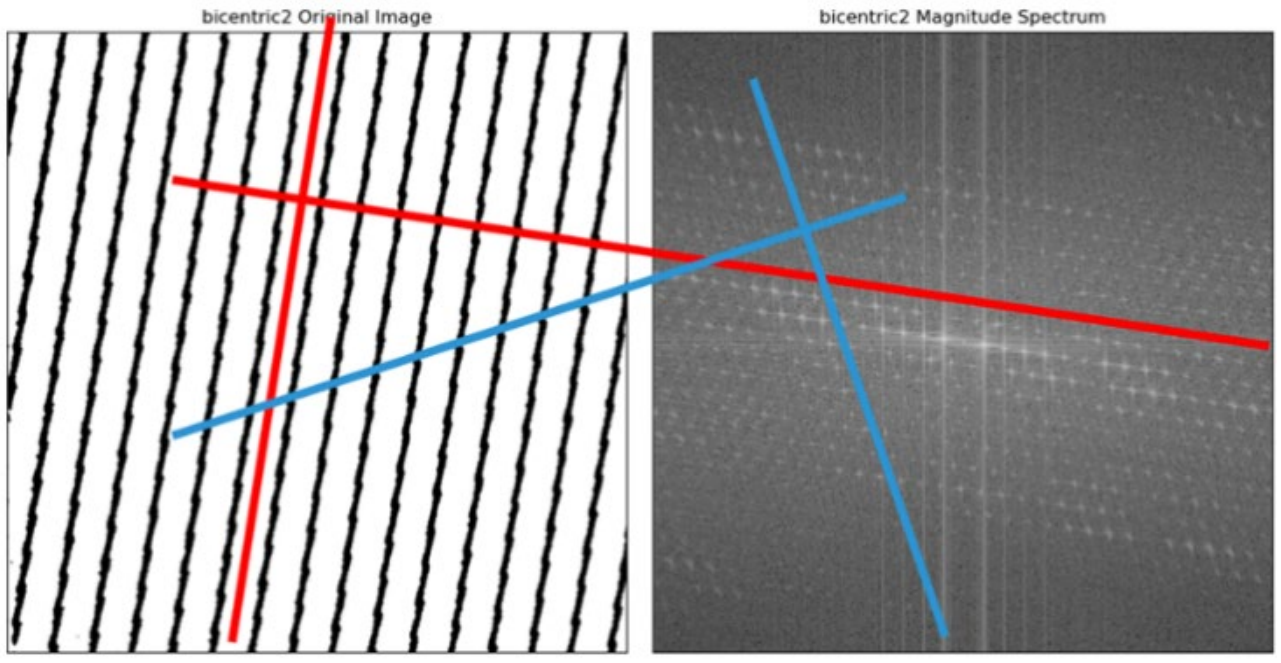

Figure 8 Spatial (a) and frequency domain (b) of degraded tilt line

Based on the direction and position of these lines, a specific region in the frequency domain is determined. A parallelogram-shaped area is defined to cover the aliasing frequencies, and frequencies with high magnitudes within this area are set to zero to effectively suppress the artifacts. In practice, during the filtering process, we preserved only the straight line oriented perpendicularly to 80 degrees, specifically the 170-degree line segment, while discarding all other frequencies.

After masking the aliasing frequencies, the modified magnitude spectrum is combined with the original phase spectrum. The image is then reconstructed in the spatial domain using the Inverse Fourier Transform. As a result of this selective filtering, the nodes that caused disturbances in the initial image were effectively removed.

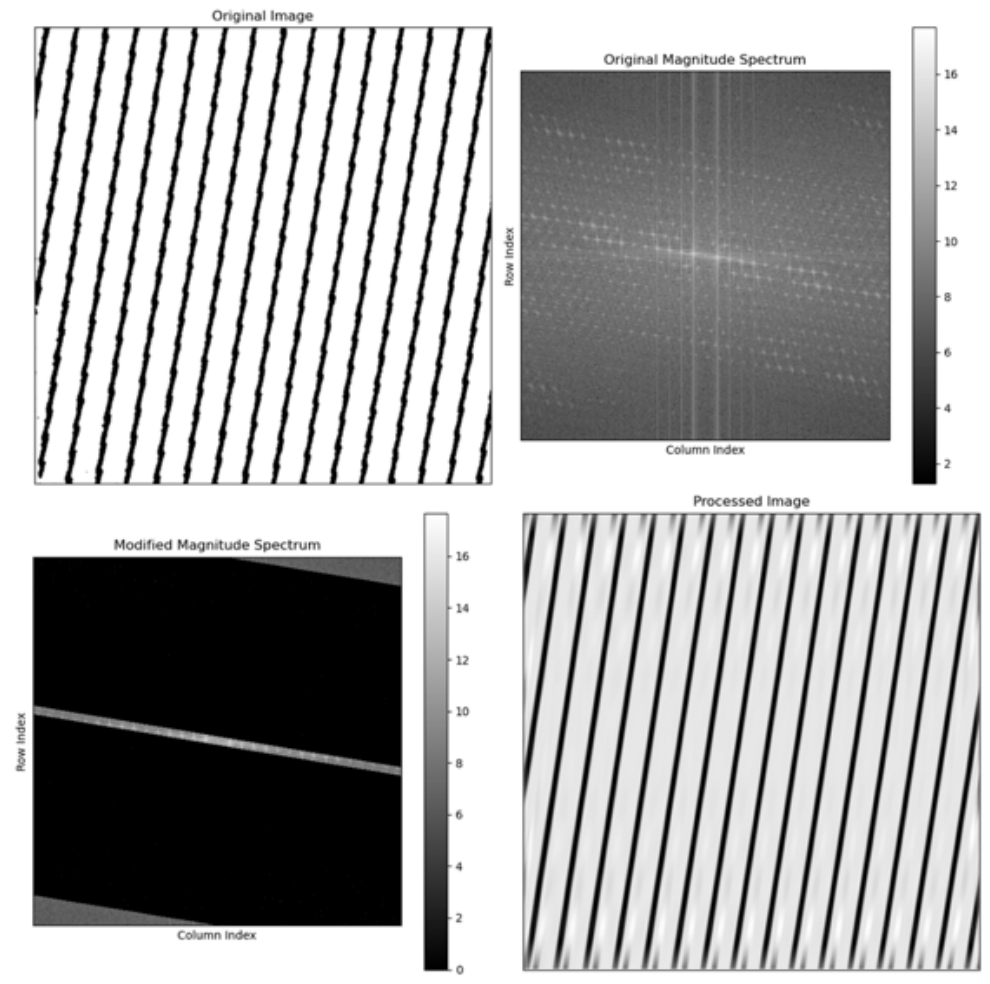

Figure 11 (a) Original image (b)Original magnitude spectrum (c) Modified magnitude spectrum (d) Processed image

To further enhance the visual quality, the reconstructed image undergoes normalization, and optional binarization is applied to highlight key features of the image while removing residual noise.

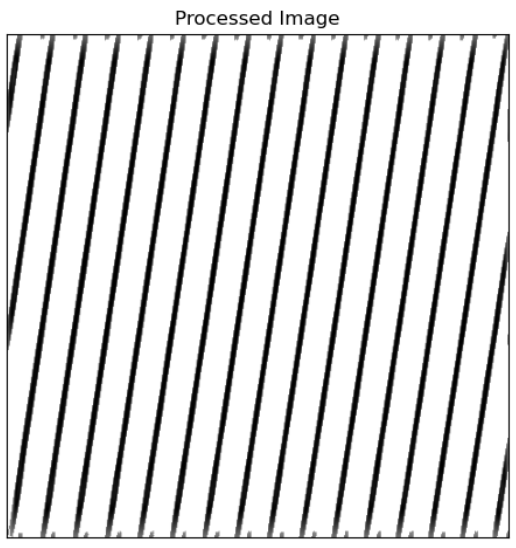

Figure 12 Binarization of processed image

This method achieves effective aliasing suppression by analyzing the frequency domain and masking specific frequencies. When comparing the original and processed images, it is evident that the aliasing frequencies are eliminated from the magnitude spectrum, resulting in improved image clarity and smoother edges, demonstrating the effectiveness of the approach.

Similarly, even if the black line assumes a greater thickness, the above methodology remains viable and effective.

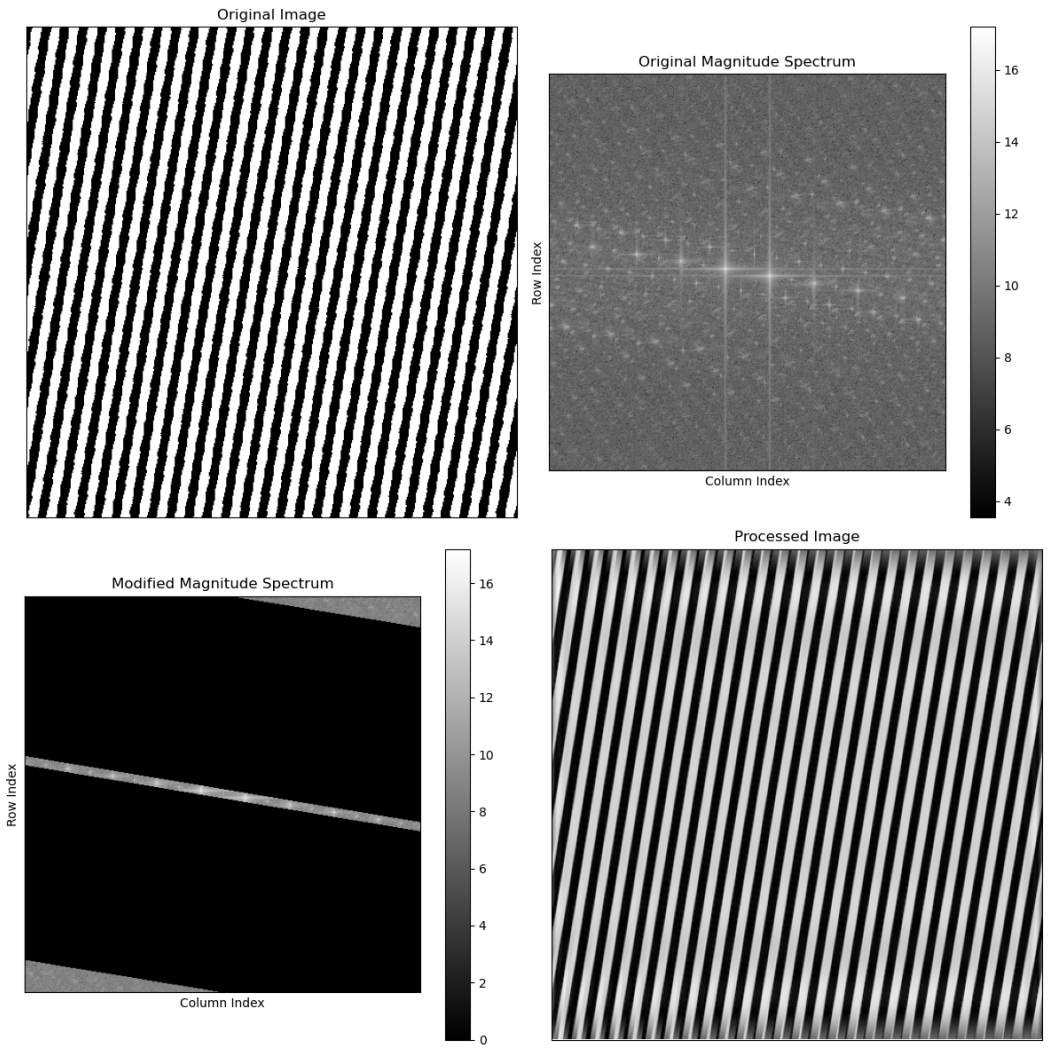

Figure 13 Processing image for a thicker line. (a) Original image (b)Original magnitude spectrum (c) Modified magnitude spectrum (d) Processed image

## IV. DEGRADATION FUNCTION FILTERING

However, using frequency filtering is not enough. The ultimate purpose of study degradation is to improve displaying performance. To achieve that, a precise degradation function is needed.

If the original image and its degraded version are available, the degradation function can be estimated by comparing the two. First, analyze the extent of blurring and noise in the degraded image, assuming the degradation is caused by the application of a degradation function to the original image, with some noise added. Using frequency domain analysis, the degradation function can be derived from the relationship between the original and degraded images. This function describes the nature of the degradation, such as the extent or direction of the blurring.

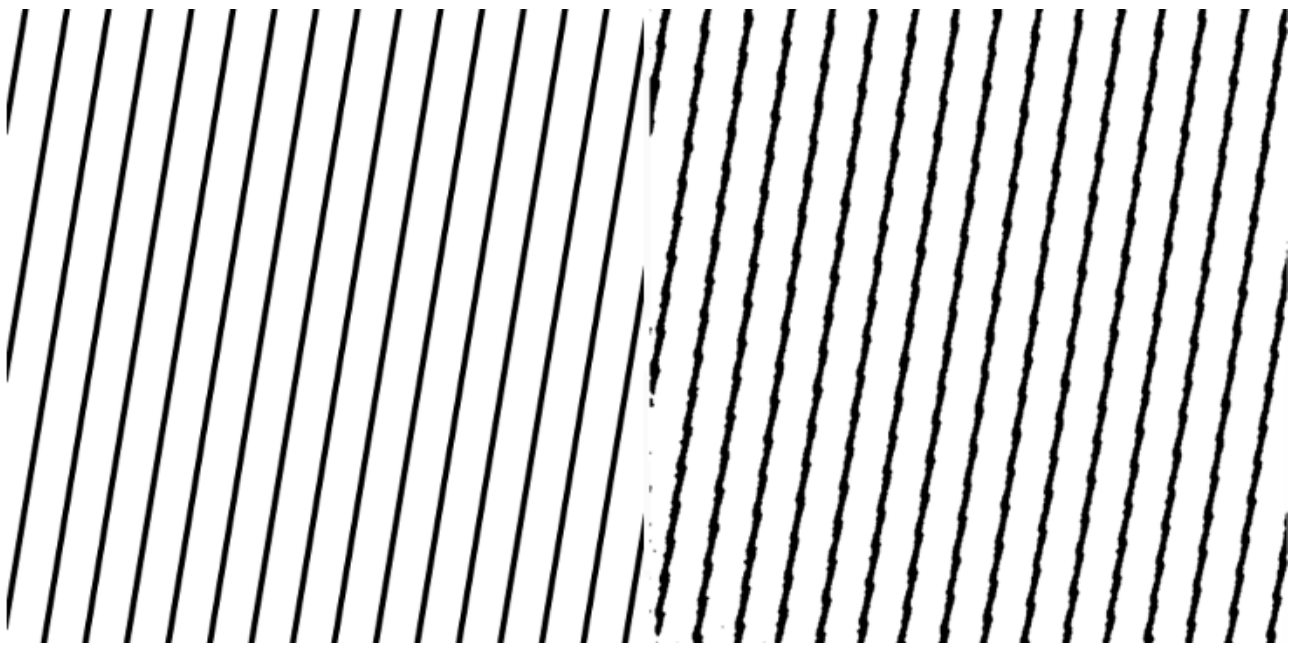

Figure 14 (a) original image and (b) the degraded image

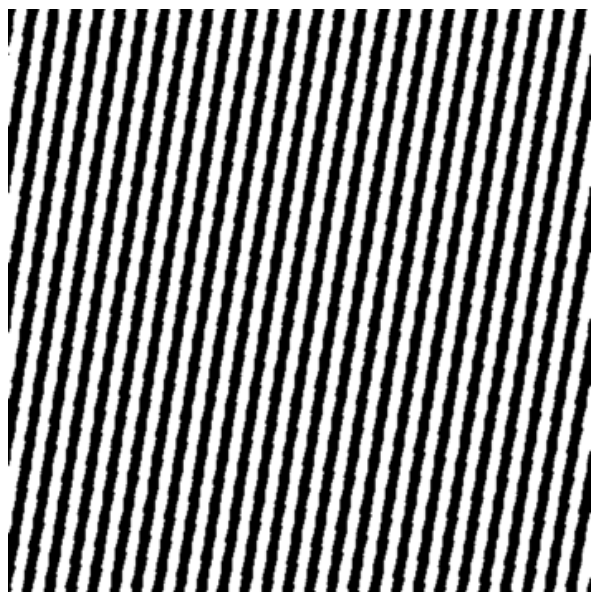

Figure 15 image to be processed

Once the degradation function is determined, it can be used to process other degraded images. For inverse filtering, the degradation function is directly inverted to restore image details. This method is highly sensitive to noise and is best suited for cases where the degradation function is well-defined and noise is minimal. By applying the inverse operation, the main structure of the image can be reconstructed, though care must be taken to avoid amplifying noise.

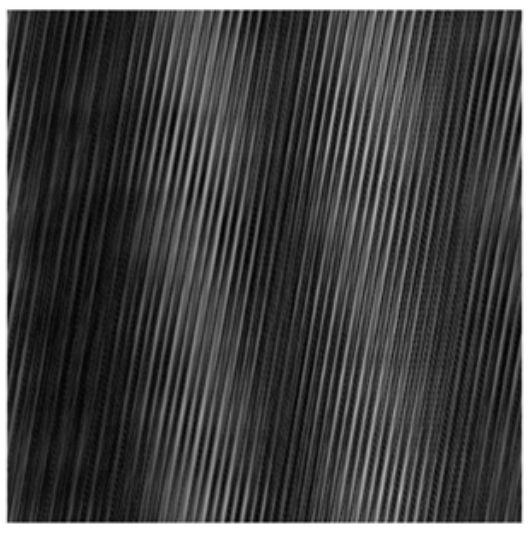

Figure 16 inverse filtering result

For radial filtering, if the degradation exhibits radial symmetry (e.g., the blur is distributed around a central point), a filter can be designed to focus on features in the frequency domain that depend on the radius. For instance, low-frequency components can be preserved to reduce high-frequency noise, or specific frequency ranges can be enhanced to restore blurred details. Radial filtering is particularly effective in handling images with radially symmetric degradation.

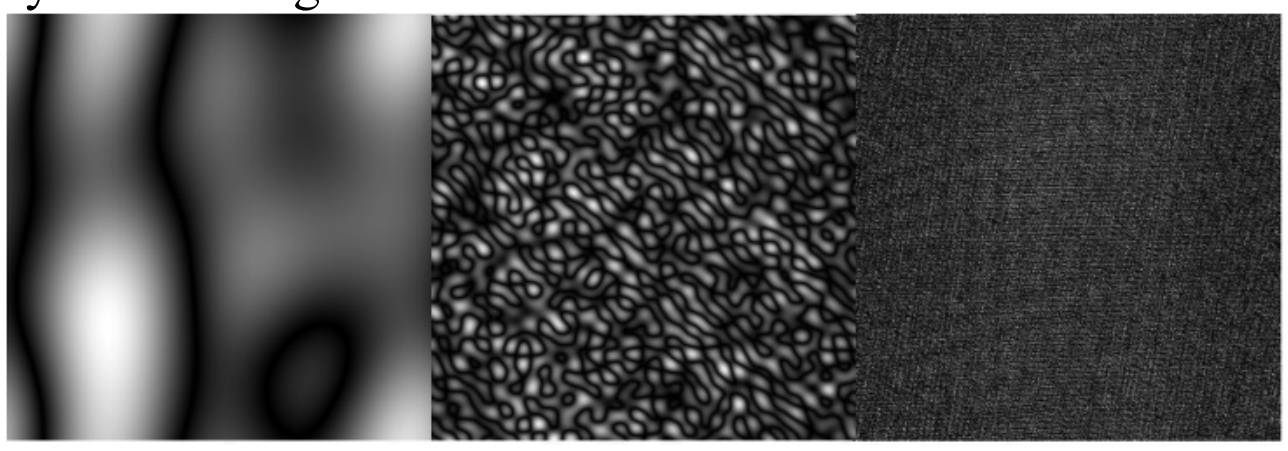

Figure 17 results with different radial restrictions

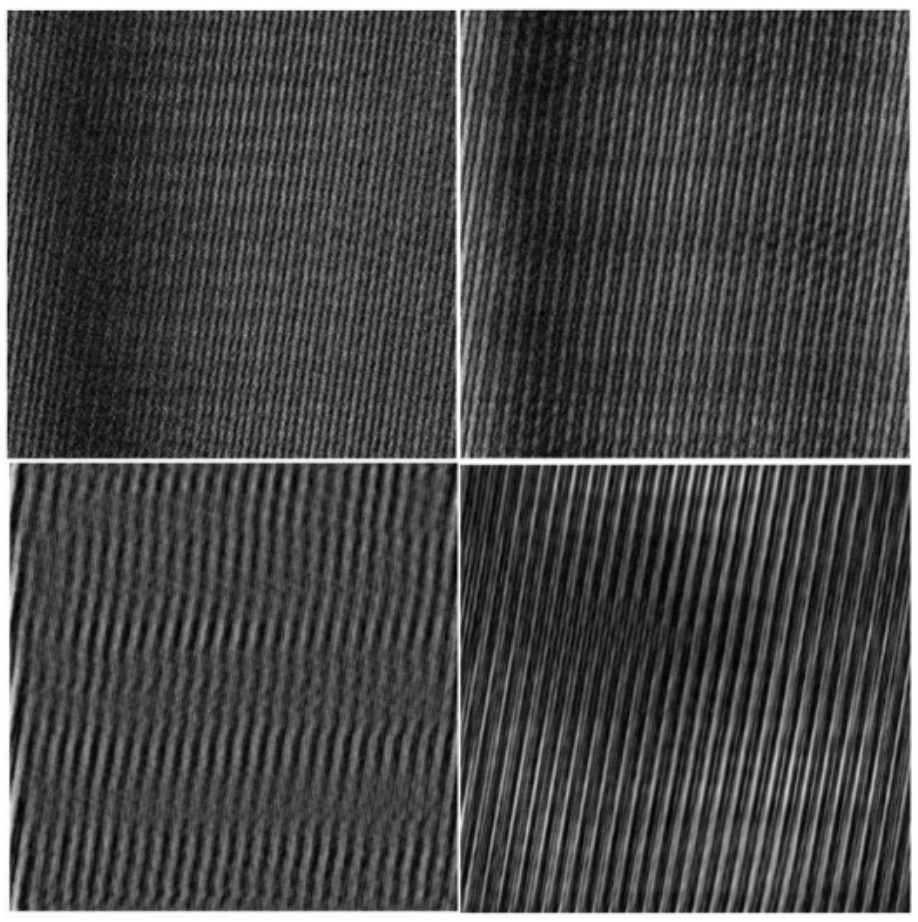

Figure 18 wiener filtering result

When noise significantly affects the image, Wiener filtering can be used. This method combines the degradation function, noise characteristics, and statistical information of the image to achieve an optimal balance between noise reduction and detail preservation. Wiener filtering is especially suitable for restoring degraded images in high-noise scenarios and can greatly enhance the quality of the restoration.

Since there is noise at high frequencies, we need to apply the radial Wiener filtering method. After setting the cutoff frequency to 17, we got the following figure.

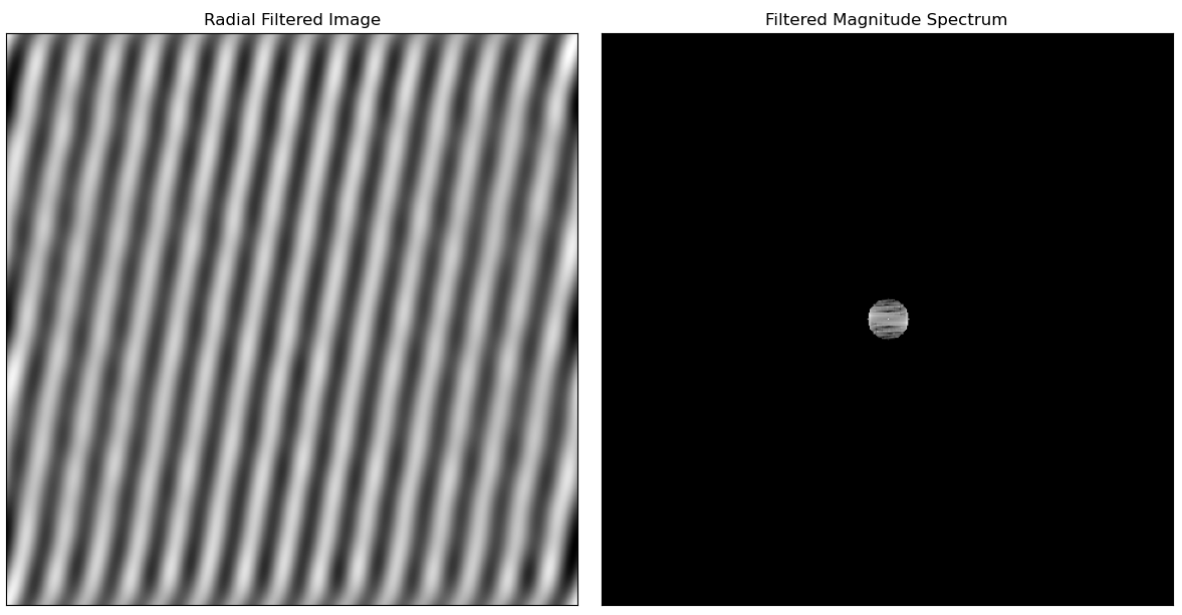

Figure 19 (a) Radial filtered image (b) Filtered magnitude sspectrum

After setting the threshold of "Custom Binarization" to 140, we could get the final image.

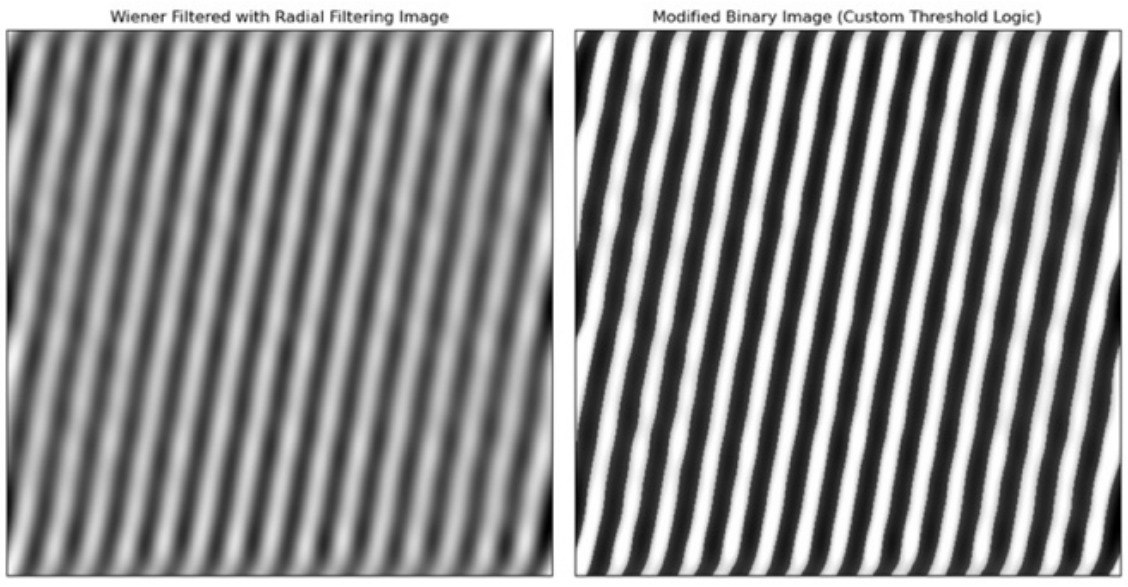

Figure 20 (a) Winer filtered with radial filtering image (b) Modified binary image (custom threshold logic)

Through the degradation function, we can see that the original straight line is thickened and blurred, thus eliminating the nodes affected by the cylindrical lens. From this, we can conclude that the degradation function and radial Wiener filtering are very effective in dealing with aliasing.

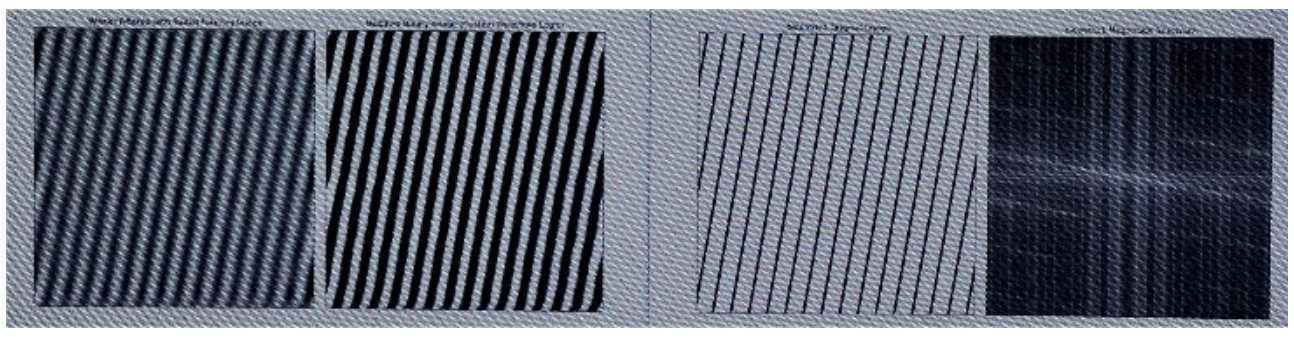

Figure 21 (a) Winer filtered with radial filtering image on 3D display device (b) Modified binary image on 3D device display (c) Original image on 3D display device (d) Frequency Spectrum of original image on 3D display device

## V. Discussion

In the subsequent restoration of degraded images, an approach can be attempted to directly process the image in the spatial domain. This involves measuring the pixel count at each node and slicing and shifting the image along the direction perpendicular to the diagonal lines based on the measurements to achieve image restoration. To further enhance the effectiveness of this method, preprocessing steps such as binarization, noise reduction, and contrast enhancement can be incorporated to improve the accuracy of node detection. Additionally, subpixel-level edge detection or frequency domain analysis can be utilized to determine node positions from both spatial and frequency perspectives. During the shifting operation, the cutting range and shifting amount can be dynamically adjusted according to the degree of degradation in different regions, avoiding errors caused by uniform global processing. Furthermore, local elastic transformations or affine transformations can be applied to correct distortions, and interpolation and filtering can be used to smooth the restored image, thereby further optimizing the final result.

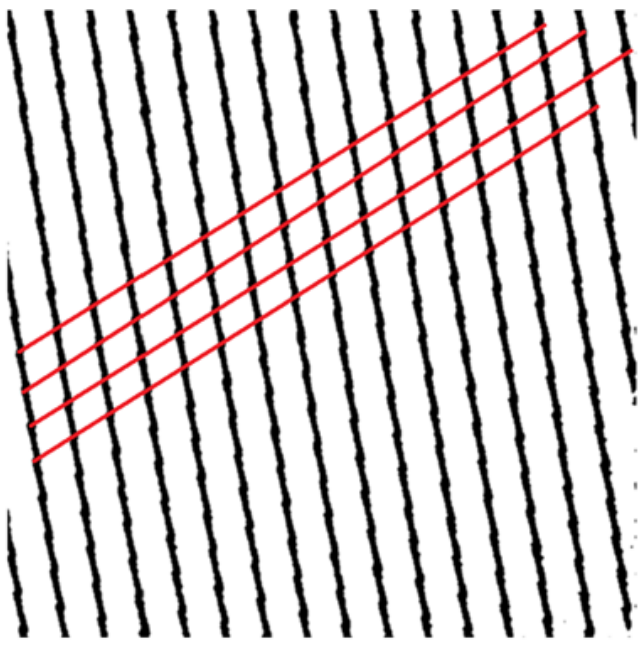

Figure 19 cutting direction and distance

Another challenging issue arises when processing text images, especially when the text is small. In such cases, the image loss caused by binarization and median filtering far exceeds the degradation caused by the cylindrical lens distortion. In other words, the losses introduced during preprocessing are significantly greater than the distortions that need to be corrected. This greatly increases the difficulty of processing such images.

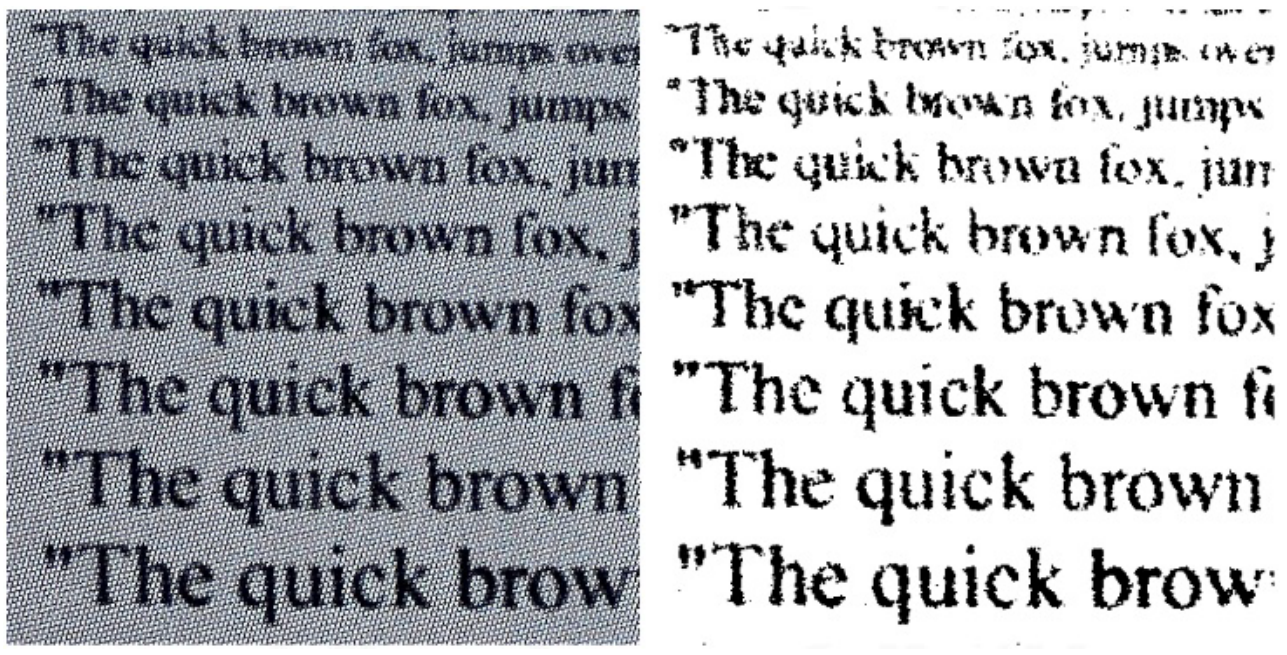

Figure 20 text images before(a)/after(b) processing

This project was developed entirely independently by the team, without borrowing any online resources or open source code, demonstrating the team's independent ability in the process of exploration and innovation.

The implementation of the project is closely integrated with the course learning, and the filtering technology and degradation function theory are cleverly used, which is highly consistent with the course content, reflecting the close connection and in-depth application of project research and course knowledge points.

## VI. Conclusion

In this work, we successfully locate the root cause of node blurring in the naked eyes 3D system. By simplified the problem into the degradation of 80 degree tilt lines, we proposed two methods that are both capable of eliminating degradation and improve displaying performance. The Frequency-domain filtering could effectively suppress the blurring effect while Radial Winer filtering can not only eliminate degeneration but also have potential to be utilized in making preprocessing image and fundamentally settle down the problem.

## ACKNOWLEDGMENT

Thanks to Professor Dong and teaching assistants for their guidance on these experimental lessons.

## CODES

### (1) Wiener Filter

```
if image1 is not None and image2 is not None:
    # 绘制频谱图
    # plot_image_and_spectrum(image1, spectrum1, "bicentric1")
    # plot_image_and_spectrum(image2, spectrum2, "bicentric2")

    # 计算频谱图之比 K = spectrum1 / spectrum2
    K = spectrum1 / (spectrum2)  # 避免除零错误
    # plt.figure(figsize=(6, 6))
    # plt.imshow(np.log(1 + K), cmap='gray')
    # plt.title('K = Spectrum1 / Spectrum2')
    # plt.xticks([]), plt.yticks([])
    # plt.colorbar()
    # plt.show()

    # 用 K 调整 bicentric2 的幅度，并保留 bicentric1 的相位
    phase1 = np.angle(fshift1)
    phase2 = np.angle(fshift2)
    phase3 = np.angle(fshift3)
    # spectrum_magnitude = np.abs(fshift3)
    spectrum_magnitude = np.abs(fshift1)

    spectrum_modified = spectrum_magnitude * K

    # 合成新的频谱：使用调整后的幅度和 bicentric1 的相位
    # fshift_modified = spectrum_modified * np.exp(1j * (2 * phase3))
    fshift_modified = spectrum_modified * np.exp(1j * (2 * phase1 - phase2))

    # 应用径向滤波
    cutoff = 17  # 截止频率
    radial_mask = radial_filter(spectrum1.shape, cutoff, filter_type="lowpass")
    filtered_spectrum = fshift_modified * radial_mask

    # 将修正后的频谱转换回图像（径向滤波后）
    f_ishift_filtered = np.fft.ifftshift(filtered_spectrum)
    image_filtered = np.abs(np.fft.ifft2(f_ishift_filtered))
```

The first part of the code primarily extracts the frequency spectrum information from the image. Through Fourier transform, the image is converted from the spatial domain to the frequency domain, and the corresponding phase information (`phase1`, `phase2`, `phase3`) and magnitude (`spectrum_magnitude`) are extracted. These serve as the foundation for subsequent operations.

In the second part, the magnitude of the spectrum is adjusted using a scaling factor `K` to generate a new magnitude `spectrum_modified`. Then, combined with the custom formula `(2 * phase1 - phase2)` and the adjusted magnitude, the code reconstructs a new spectrum `fshift_modified`. This operation aims to integrate the information from two images.

The third part applies a radial low-pass filter to the new spectrum. The radial filter, based on the set cutoff frequency `cutoff`, creates a mask `radial_mask` to retain the low-frequency components of the spectrum while removing high-frequency noise. The filtered spectrum is stored in `filtered_spectrum`.

The final step restores the filtered spectrum back to the spatial domain using the inverse Fourier transform. This process includes reversing the spectrum’ s central shift, performing the inverse transform, and taking the absolute value to generate the final filtered image `image_filtered`, which can be used for further visualization or processing.

Pseudocodes:

Step 1: Input the image and compute Fourier transform
image = input_image()
f_transform = fft2(image)
phase1, phase2, phase3 = extract_phases(f_transform)
spectrum_magnitude = extract_magnitude(f_transform)
Step 2: Calculate the frequency scaling factor K and adjust magnitude
K = calculate_scaling_factor()
spectrum_modified = spectrum_magnitude * K
Step 3: Recombine spectrum using adjusted magnitude and custom phases
fshift_modified = spectrum_modified * exp(1j * (2 * phase1 - phase2))
Step 4: Generate radial low-pass filter and apply it
radial_mask = generate_radial_lowpass_filter()
filtered_spectrum = fshift_modified * radial_mask
Step 5: Perform inverse Fourier transform to obtain the processed image
image_filtered = ifft2(filtered_spectrum)
Step 6: Output the filtered image
output_image(image_filtered)

### (2) Direct Filter

Pseudocodes:

Import the necessary libraries for image processing and mathematical operations. Define a function named modify_parallelogram that takes in the magnitude spectrum of an image and the coordinates of the parallelogram's four vertices. This function calculates the boundaries of the parallelogram based on the provided coordinates and iterates over the pixels within this region. If a pixel falls within the defined region and meets a specific condition, its value in the magnitude spectrum is set to zero.

Read the grayscale image from the file and compute its Fast Fourier Transform (FFT). From the FFT result, extract the magnitude and phase spectra. The magnitude spectrum contains information about the strength of different frequency components, while the phase spectrum holds information about their position in the frequency domain.
Define the four vertices of the parallelogram and use the modify_parallelogram function to apply the changes to the magnitude spectrum based on the parallelogram ’ s boundaries. After the modification, apply logarithmic scaling to the magnitude spectrum in order to avoid very small values that could cause the image to appear entirely black.

After processing the magnitude spectrum, combine it with the original phase spectrum to form the modified frequency domain data. Perform an inverse Fast Fourier Transform (IFFT) on this modified data to convert it back into the spatial domain and reconstruct the processed image. Use Matplotlib to display the original image, the original magnitude spectrum, the modified magnitude spectrum, and the processed image.

Normalize the processed image to ensure the pixel values are in the correct range for display. Then, implement custom binarization by iterating over the image’ s pixels. For any pixel whose value exceeds a given threshold, set it to white. Finally, display the binarized version of the processed image using Matplotlib.

This code performs image processing in the frequency domain, demonstrating how modifying the frequency spectrum affects the image through Fourier transforms and inverse Fourier transforms.

Code explanation:

First, the code imports the necessary libraries: `numpy` for mathematical operations, `matplotlib.pyplot` for displaying images, and `cv2` for reading and processing images. It then defines a function `modify_parallelogram`, which takes the magnitude spectrum and four vertex coordinates as inputs. The function calculates the boundaries of a parallelogram region and iterates over the pixels within this region. If a pixel meets certain conditions, its corresponding frequency component is set to zero, removing specific frequencies from the spectrum.

Next, the code reads an image and converts it to grayscale. Grayscale images are commonly used in image processing because they simplify the data and reduce computational complexity. Then, the code performs a Fast Fourier Transform (FFT) on the image, converting it from the spatial domain to the frequency domain and obtaining the magnitude and phase spectra. The magnitude spectrum represents the strength of the frequency components, while the phase spectrum provides the position and phase information of these frequencies.

After obtaining the frequency spectrum, the code defines a parallelogram region using the four predefined vertices. The `modify_parallelogram` function is called to modify the frequency components within this region. Specifically, the function iterates through the pixels inside the parallelogram and sets the spectrum values to zero if they meet certain criteria, removing unwanted frequency components from the image. This modification is often used to eliminate noise or unnecessary frequencies in an image.

Once the spectrum is modified, the code combines the modified magnitude spectrum with the original phase spectrum to generate new frequency-domain data. It then performs an inverse Fast Fourier Transform (IFFT) to convert the frequency-domain data back into the spatial domain, producing the processed image. This step allows the effect of the modified frequency spectrum to be reflected in the image.

After the image is obtained through the inverse Fourier transform, the code normalizes the image. This step ensures that the pixel values fall within a suitable range for display, avoiding display issues. The normalization adjusts the image's pixel values to a range between 0 and 255. Following this, the code performs a thresholding operation, turning pixels with values greater than a specified threshold white, while leaving others black. This binarization technique is commonly used for image segmentation or to highlight certain regions of interest.

Finally, the code uses `matplotlib` to display four images: the original image, the original magnitude spectrum, the modified magnitude spectrum, and the processed image. These images allow us to visually observe the changes in the image after Fourier transform and frequency spectrum modification, helping to understand how changes in the frequency domain affect the spatial domain representation of the image.

```
def modify_parallelogram(magnitude_spectrum, points):
    # [illegible]
    top_left = points[0]
    top_right = points[1]
    bottom_left = points[2]
    bottom_right = points[3]
    rows, cols = magnitude_spectrum.shape
    crow, ccol = rows // 2, cols // 2

    # [illegible]
    left_slope = (bottom_left[1] - top_left[1]) / (bottom_left[0] - top_left[0])
    right_slope = (bottom_right[1] - top_right[1]) / (bottom_right[0] - top_right[0])

    # [illegible]
    for x in range(top_left[0], bottom_left[0] + 1):
        # [illegible]
        top_y = int(top_left[1] + left_slope * (x - top_left[0]))
        bottom_y = int(top_right[1] + right_slope * (x - top_right[0]))

        # [illegible]
        if top_y > bottom_y:
            top_y, bottom_y = bottom_y, top_y

         # [illegible]
        for y in range(top_y, bottom_y + 1):
            # [illegible]
            if 0 <= y < magnitude_spectrum.shape[0] and 0 <= x < magnitude_spectrum.shape[1]:
                # if y != int(- slope * (x-ccol) + crow) and y != int(- slope * (x-ccol+1) + crow) and y != int(- slope * (x-ccol+2) + crow)\
                #     and y != int(- slope * (x-ccol+3) + crow) and y != int(- slope * (x-ccol+4) + crow) and y != int(- slope * (x-ccol+5) + crow):
                #     if magnitude_spectrum[y, x] > 5000:
                #         magnitude_spectrum[y, x] = 0
                if all(y != int(-slope * (x - ccol + i) + crow) for i in range(-30, 31)):
                    if magnitude_spectrum[y, x] > 60:
                        magnitude_spectrum[y, x] = 0


    return magnitude_spectrum
```

(3)Preprocessing

Binarize:

This code uses the OpenCV library for image processing, with the primary function of binarizing the input image and saving the result as a new image file. First, it imports the necessary libraries: cv2 for image processing, numpy for array operations, and os for handling file paths and other operations.

The binarize_image function performs the binarization process. It takes the image path as input and loads the image in grayscale mode using cv2.imread. If the image fails to load, it raises an error. Then, it uses cv2.threshold to set the pixel values below a specified threshold (default is 20) to 0, and all other pixel values to 255, completing the binarization process.

Next is the save_image function, responsible for saving the processed binary image to a specified path. It writes the image to the file using cv2.imwrite.

In the main function, the input and output image paths are first set. The binarize_image function is then called to binarize the image, and the result is stored in the binary_image variable. After that, the save_image function is called to save the binarized image to the specified location on the disk. Finally, cv2.imshow is used to display the processed image, and cv2.waitKey waits for the user to press any key before closing the display window.

The execution flow of this code is: reading the image, binarizing it, saving it, and displaying the result.

```python
def binarize_image(image_path, threshold=20):
    """根据阈值进行二值化处理"""
    # 加载图像
    image = cv2.imread(image_path, cv2.IMREAD_GRAYSCALE)

    if image is None:
        raise ValueError(f"Image not found at path: {image_path}")

    # 二值化处理：灰度值小于 threshold 的部分设为 0，其他设为 255
    _, binary_image = cv2.threshold(image, threshold, 255, cv2.THRESH_BINARY)

    return binary_image

def save_image(image, save_path):
    """保存处理后的图像"""
    cv2.imwrite(save_path, image)

def main():
    # 设置图像路径和保存路径
    input_image_path = 'res/degraded.png'  # 输入图像路径
    output_image_path = 'binary/binary_output_20.png'  # 输出图像路径

    # 二值化处理
    binary_image = binarize_image(input_image_path)

    # 保存二值化后的图像
    save_image(binary_image, output_image_path)

    # 显示图像
    cv2.imshow("Binarized Image", binary_image)
    cv2.waitKey(0)
    cv2.destroyAllWindows()
```

Median filter:

This code is designed for batch processing images using different types of median filtering techniques, including single median filtering, double median filtering, and adaptive median filtering.

The process_images_median function handles applying a single median filter to each image in a list of image paths. Similarly, process_images_double_median applies double median filtering, and process_images_adaptive_median uses a dap tive median filtering. Each of these functions processes the images by calling the corresponding filtering function (such as process_median_filter, process_double_median_filter, or process_adaptive_median_filter), and then saves the results to a specified output directory.

In the main section of the code, a list of image paths is defined, and the various filtering methods are applied. For example, a kernel size of 5 is used for the median filters, and the maximum size for the adaptive median filter is set to 7. The filtered images are then saved to an "output" directory. Additionally, a noise variance value (`V_noise`) is included, but it doesn't appear to be used directly in the code, possibly for future enhancements.

This structure allows for efficient batch processing of multiple images using different filtering techniques, with the results saved in a specified location.

```python
def process_images_median(image_paths, kernel_dim, output_dir):
    """批量处理多张图片，一次中值滤波"""
    for image_path in image_paths:
        process_median_filter(image_path, kernel_dim, output_dir)

# 批量处理图像（双重中值滤波）
def process_images_double_median(image_paths, kernel_dim, output_dir):
    """批量处理多张图片，两次中值滤波"""
    for image_path in image_paths:
        process_double_median_filter(image_path, kernel_dim, output_dir)

# 批量处理图像（自适应中值滤波）
def process_images_adaptive_median(image_paths, s_max, output_dir):
    """批量处理多张图片，自适应中值滤波"""
    for image_path in image_paths:
        process_adaptive_median_filter(image_path, s_max, output_dir)
```